\documentclass{aa}  
\usepackage{graphicx}
\usepackage{txfonts}
\usepackage[urlcolor=blue]{hyperref}
\usepackage{breakurl}
\newcommand{\lya}{Ly$\alpha$}
\newcommand{\lyb}{Ly$\beta$}
\newcommand{\lyc}{Ly$\gamma$}
\newcommand{\ha}{H$\alpha$}

\titlerunning{Broadband \lya\ Observations During Solar Flares From SDO/EVE}

\authorrunning{Milligan \& Chamberlin}

\begin{document}

\title{The Anomalous Temporal Behaviour of Broadband \lya\ Observations During Solar Flares From SDO/EVE}

\author{Ryan O. Milligan\inst{1,2,3} \& Phillip C. Chamberlin\inst{2}}

\institute{Astrophysics Research Centre, School of Mathematics \& Physics, Queen's University Belfast, University Road, Belfast, Northern Ireland, BT7 1NN \\ \email{r.milligan@qub.ac.uk} \and Solar Physics Laboratory (Code 671), Heliophysics Science Division, NASA Goddard Space Flight Center, Greenbelt, MD 20771, USA \and Department of Physics, Catholic University of America, 620 Michigan Avenue, Northeast, Washington, DC 20064, USA}

\date{Received Month Day, Year; accepted Month Day, Year}


\abstract
{Despite being the most prominent emission line in the solar spectrum, there has been a notable lack of studies devoted to variations in \lya\ emission during solar flares in recent years. However, the few examples that do exist have shown \lya\ emission to be a substantial radiator of the total energy budget of solar flares (on the order of 10\%). It is also a known driver of fluctuations in earth's ionosphere. The EUV Variability Experiment (EVE) onboard the Solar Dynamics Observatory now provides broadband, photometric \lya\ data at 10~s cadence with its Multiple EUV Grating Spectrograph-Photometer (MEGS-P) component, and has observed scores of solar flares in the 5 years since it was launched. However, the MEGS-P time profiles appear to display a rise time of tens of minutes around the time of the flare onset. This is in stark contrast to the rapid, impulsive increase observed in other intrinsically chromospheric features (\ha, \lyb, LyC, C~III, etc.). Furthermore, the emission detected by MEGS-P peaks around the time of the peak of thermal soft X-ray emission, rather than during the impulsive phase when energy deposition in the chromosphere - often assumed to be in the form of nonthermal electrons - is greatest. The time derivative of \lya\ lightcurves also appears to resemble that of the time derivative of soft X-rays, reminiscent of the Neupert Effect. Given that spectrally-resolved \lya\ observations during flares from SORCE/SOLSTICE peak during the impulsive phase as expected, this suggests that the atypical behaviour of MEGS-P data is a manifestation of the broadband nature of the observations. This could imply that other lines and/or continuum emission that becomes enhanced during flares could be contributing to the passband. Users are hereby urged to exercise caution when interpreting broadband \lya\ observations of solar flares. Comparisons have also been made with other broadband \lya\ photometers such as PROBA2/LYRA and GOES/EUVS-E.}


\maketitle

\section{Introduction}
\label{intro}

The Lyman-alpha (\lya; 1216\AA) transition (2p--1s) of hydrogen results in the strongest emission line in the solar spectrum. It is an optically-thick line formed in the mid-to-upper chromosphere and recent studies have suggested that it is responsible for radiating a significant fraction of the nonthermal energy deposited in the chromosphere during solar flares ($\sim$10\%; \citealt{nusi06,rubi09,mill14}). \lya\ is also known to be a driver of changes in terrestrial ionospheric density in the D-layer (80--100~km; \citealt{tobi00}) in conjunction with soft X-rays (SXR), although the \lya\ contribution is smaller for flares that occur closer to the solar limb due to absorption along the line of sight \citep{wood06,qian10}. Understanding variations in \lya\ are therefore a major priority for both flare physics and space weather research. However, despite the importance of \lya\ as a solar diagnostic there are relatively few papers in the literature that discuss changes in \lya\ emission during solar flares. (For a recent review of \lya\ and other chromospheric EUV flare observations, see \citealt{mill15}.) This paper aims to highlight the inconsistencies between currently available \lya\ datasets.

Early irradiance measurements in \lya\ were routinely carried out by instruments such as the Solar Stellar Irradiance Comparison Experiment (SOLSTICE) onboard the Solar Radiation and Climate Experiment (SORCE), and the Solar EUV Experiment (SEE) onboard Thermosphere Ionosphere Mesosphere Energetics and Dynamics (TIMED) satellite. These, and other, instruments often only had a duty cycle of a few per cent so flares were either missed or averaged over. One notable exception was during the Halloween flares of 2003 when spectrally-resolved \lya\ line profiles were obtained at high cadence using SORCE/SOLSTICE (\citealt{wood04}; see Section~\ref{sec:sorce_lya}). In 2010, NASA launched the Solar Dynamics Observatory (SDO; \citealt{pesn12}) as part of its Living With A Star program to help understand the effects of solar EUV variability on the earth. One of the three instruments onboard is the EUV Variability Experiment (EVE; \citealt{wood12}). Its prime objective is to improve upon previous measurements by recording changes in the solar EUV irradiance on flaring timescales (10~s). It achieves this through its MEGS-A (Multiple EUV Grating Spectrograph) and -B components, which cover the 60--370\AA, and 370--1050\AA\ wavelength ranges respectively. It also comprises a 106\AA\ wide, broadband photometric diode, MEGS-P, which is placed at the minus first order of the MEGS-B grating centred on the \lya\ line.

Flare-related enhancements in \lya\ from EVE data were first reported by \cite{mill12} for the 15 February 2011 X2.2 flare, and a follow-up study claimed that \lya\ emission made up 6-8\% of the total measured radiated losses from the chromosphere for that event \citep{mill14}. In contrast, \cite{kret13} found only a 0.6\% increase in \lya\ emission during an M2 flare using the LYRA radiometer (1200-1230\AA; \citealt{benm09}) onboard PROBA2. More strikingly though was that in both cases the temporal behaviour of \lya\ appeared to mimic that of the SXR emission, rather than of the impulsive hard X-ray (HXR) emission as one might expect for intrinsic chromospheric emission. \cite{kret13} also found that by taking the time derivative of the \lya\ profile, the resulting peaks closely resembled those of the derivative of the SXR lightcurve. The authors cited this as evidence for pre-flare heating prior to the onset of accelerated particles. While the possibility of gradual heating by thermal conduction cannot be ruled out, it is more likely that other instrumental factors might have been contributing to this anomalous behaviour. This research note aims to highlight the inconsistencies between modern \lya flare observations, and to urge users to exercise caution when interpreting data from EVE and other broadband photometric instruments. The data from MEGS-P are the focus of this paper and shall be discussed in greater detail in Section~\ref{sec:eve_lya}. Spectrally-resolved \lya\ flare observations from SORCE/SOLSTICE shall be presented in Section~\ref{sec:sorce_lya} as a benchmark, while a comparison between EVE and GOES/EUVS-E \lya\ measurements is given in Section~\ref{sec:goes_lya}. A summary and conclusions shall be given in Section~\ref{sec:conc}.

\begin{figure}[!t]
\begin{center}
\includegraphics[width=0.49\textwidth]{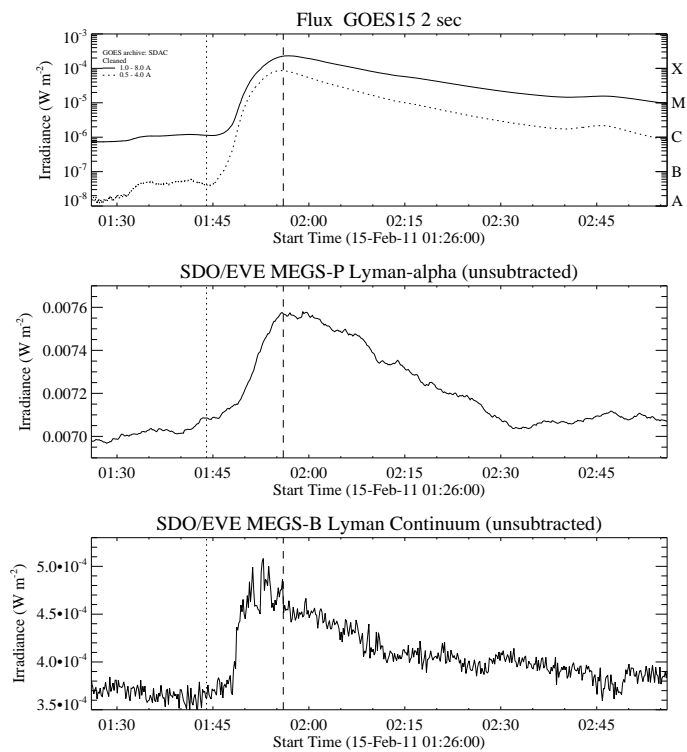}
\caption{Top panel: GOES SXR lightcurves during the X2.2 flare that occurred on 15 February 2011 in 1--8\AA\ (solid curve) and 0.5--4\AA\ (dotted curve). Middle panel: \lya\ lightcurve taken using SDO/EVE MEGS-P for the same event. Bottom panel: LyC lightcurve from SDO/EVE MEGS-B data. The vertical dotted and dashed lines in each panel mark the start and peak times of the GOES event, respectively. From \cite{mill15}.}
\label{eve_lyman_ltc}
\end{center}
\end{figure}

\section{SDO/EVE MEGS-P Flare Observations}
\label{sec:eve_lya}

SDO/EVE now provides routine, broadband ($\approx$100\AA; \citealt{hock12}) Sun-as-a-star observations in \lya\ through its MEGS-P diode at 10~s cadence, albeit with a reduced duty cycle. Until recently the MEGS-B and -P components were often only exposed to the Sun for 4 hours per day due to unforeseen instrumental degradation, although earlier in the SDO mission 24--48 hour MEGS-B flare campaigns were run during periods of very high solar activity. In October 2015, the EVE flight software was updated so that MEGS-B and -P will now respond autonomously to M-class flares or greater based on the SXR flux detected by the ESP 1--7\AA\ channel. Two 3-hour campaigns are permitted each day, in addition to 5 minutes of data taken every hour. Since the loss of MEGS-A and SAM in May 2014, the decision to only expose MEGS-B and -P during periods of increased activity, essentially makes EVE a dedicated flare instrument. Although the primary motivation behind this is to be able to quantify the total EUV flux, including \lya, incident on the earth's atmosphere during the largest solar flares for inclusion in atmospheric models, these data are also useful in terms of quantifying the composition and energy budget of radiative losses in the chromosphere.

However, in order to utilise EVE MEGS-P data\footnote{Version 5 of EVE data were used in the preparation of this paper.} correctly, it is crucial to understand what it is depicting. Figure~\ref{eve_lyman_ltc} shows the SXR (top; from the X-Ray Sensor (XRS) onboard the Geostationary Orbiting Environmental Satellites; GOES), \lya\ (middle), and the free-bound Lyman continuum\footnote{The Lyman continuum was fit over the 800--912\AA\ range in EVE MEGS-B data using techniques described in \cite{mill14}.} (LyC, with a recombination edge at 912\AA; bottom) time profiles for the 15 February 2011 X-class flare (SOL2011-02-15T01:44). \lya\ and LyC both result from transitions to the ground state of neutral hydrogen, albeit from different upper levels. However, the lightcurves of \lya\ appear to show a gradual, slowly-rising, ``GOES-like'' profile, with a rise time of 10--20 minutes, akin to that seen in SXR. (A similar behaviour was found by \citealt{kret13} using LYRA data.) The LyC profile, on the other hand, appears impulsive and bursty with a rise time of just a few minutes, and peaks during the rise in SXR. The latter case is what one might expect for a chromospheric plasma heated via Coulomb collisions (e.g., \citealt{brow71}). So why do these two profiles behave so differently given that they are formed in the same layer of the Sun's atmosphere? Is the peculiar behaviour of \lya\ merely due to the broad response function of MEGS-P, or is it indicative of how \lya\ responds to heating during large explosive events? Possible instrumental issues may include contamination of the passband by lines and/or continua from other ion species, or the core and the wings of the \lya\ line itself - which are formed at distinctly different depths in the solar atmosphere - responding to different heating mechanisms, or on different timescales. In these cases the broadband nature of the observations would therefore appear to `smooth out' the time profiles. If it is a genuine solar effect, then maybe the excitation/decay timescales for the 2p-1s transition of neutral hydrogen are longer than for other transitions or elements; or perhaps the opacity of the line is causing photons from the core to be scattered into the wings; or the \lya\ emission is actually peaking as the plasma cools rather than as it is being heated.

\begin{figure}[!t]
\begin{center}
\includegraphics[width=0.49\textwidth]{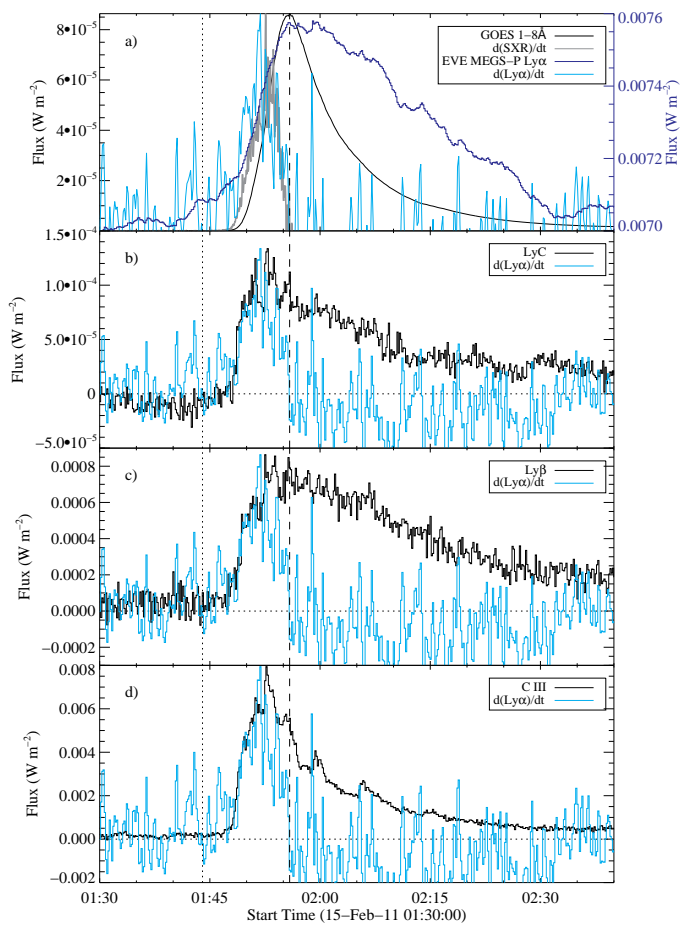}
\caption{a): A plot of \lya\ (blue) and SXR (black) emission and their derivatives (cyan and grey, respectively) during the 15 February 2011 flare. b) LyC lightcurve for the same event (black). Overplotted in cyan is the time derivative of the \lya\ lightcurve normalised to LyC. c) Lightcurve of \lyb\ (1026\AA; black curve). Overplotted in cyan is the time derivative of the \lya\ lightcurve normalised to \lyb. d) Lightcurve of C III (977\AA; black curve). Overplotted in cyan is the time derivative of the \lya\ lightcurve normalised to C III. The vertical dotted and dashed lines in each panel mark the start and peak times of the GOES event, respectively.}
\label{eve_dlyadt}
\end{center}
\end{figure}

Figure~\ref{eve_lyman_ltc} shows that there is a distinct difference between the temporal behaviour of \lya\ and LyC. However, by taking the time derivative of the \lya\ profile and normalising it to LyC there appears to be a remarkable agreement, at least on the rise phase of the 15 February 2011 flare as shown in Figure~\ref{eve_dlyadt}b. Both the LyC time profiles and the derivative of \lya\ now peak in concert with the derivative of the SXR emission, which is often taken as a proxy for the hard X-ray (HXR) emission under the assumption of the Neupert Effect \citep{neup68}. This is also in agreement with \cite{kret13}. This empirical relationship assumes that the energetic electrons that generate HXR emission through the thick-target bremsstrahlung process, are responsible for the heating and mass supply (through chromospheric evaporation) of the SXR emitted by the hot coronal plasma. In other words, SXR are an {\it indirect} consequence of the initial chromospheric heating. The fact that the \lya\ profiles exhibit a similar behaviour might suggest that the 2p--1s transition of H~I is not directly affected by collisional excitation, but perhaps by an alternative mechanism such as thermal conduction. 

The correlation between $d($\lya$)/dt$ and LyC is not unique. In Figures~\ref{eve_dlyadt}c and \ref{eve_dlyadt}d, the time derivative of \lya\ is also normalised to \lyb\ (1026\AA) and C~III (977\AA), respectively; two arbitrarily chosen, intrinsically chromospheric emission lines from MEGS-B data, and the same agreement still holds. In some other flares, however, the derivative of \lya\ precedes the LyC (or \lyb\ or C~III) emission by 1--2 minutes. The reason for this is unclear. 

\begin{figure}[!t]
\begin{center}
\includegraphics[width=0.49\textwidth]{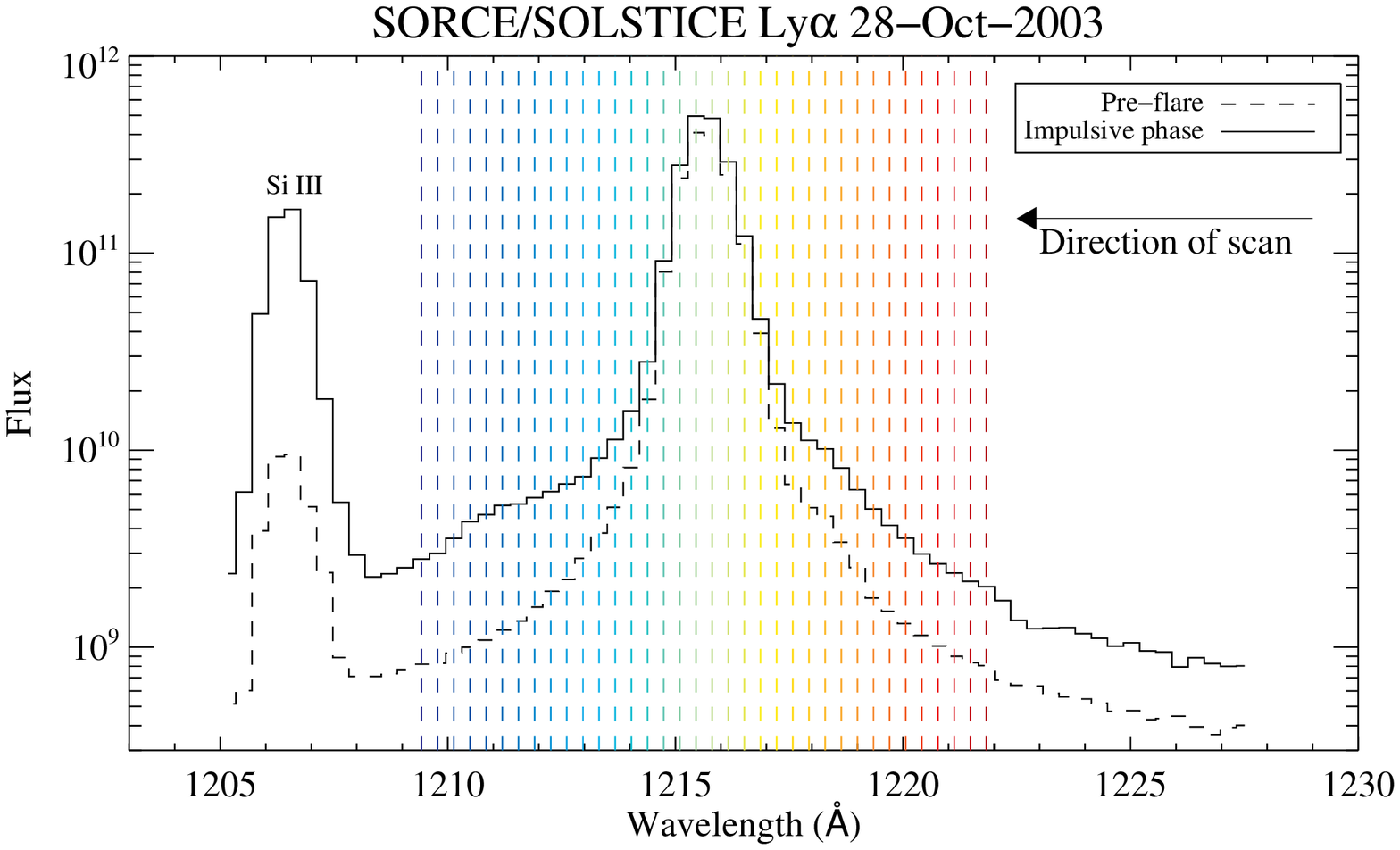}
\includegraphics[width=0.49\textwidth]{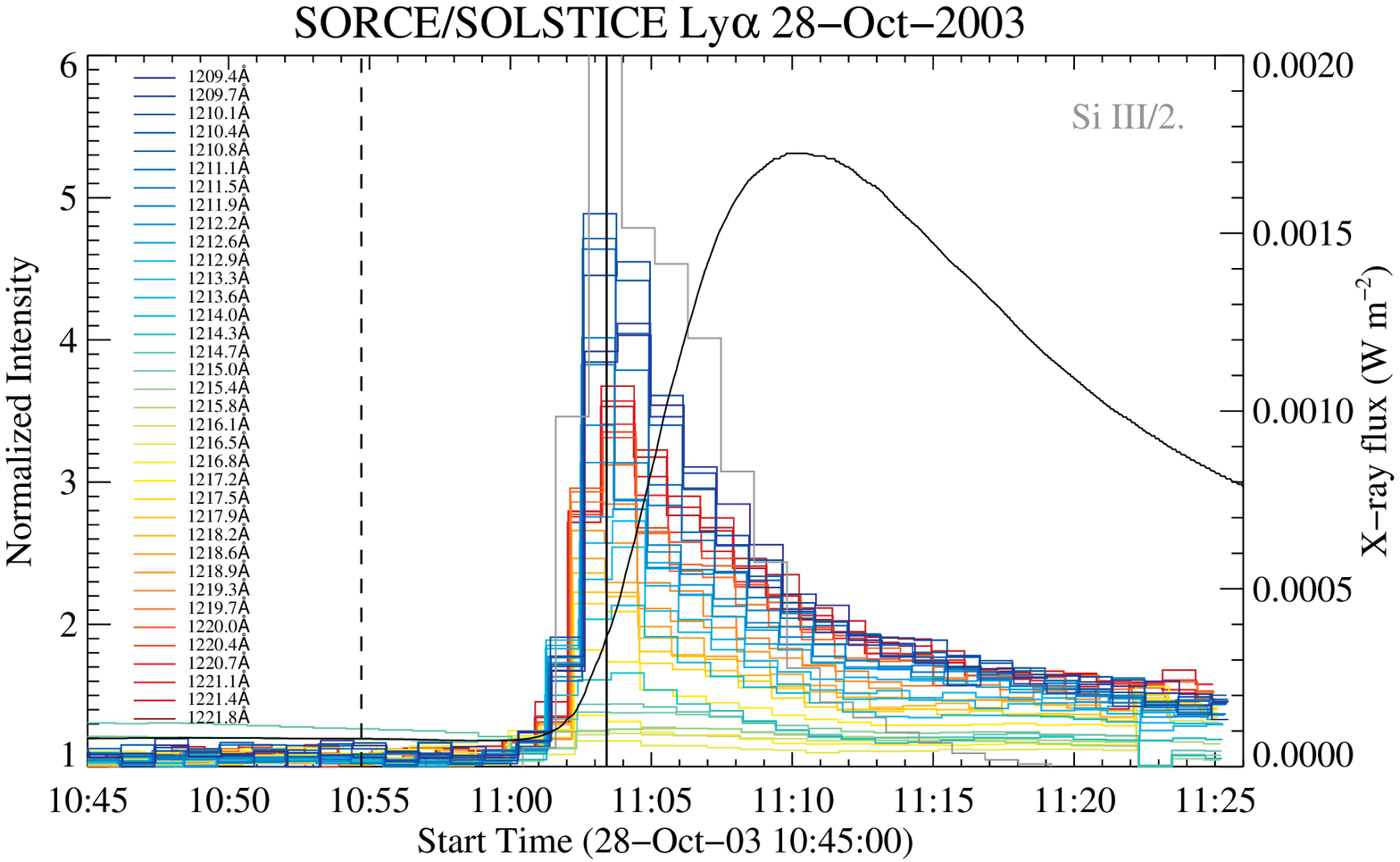}
\caption{Top panel: The \lya\ line profile observed by SORCE/SOLSTICE before (dashed curve) and during (solid curve) the 28 October 2003 flare. Also visible is the nearby Si III line. The vertical coloured lines denote the wavelengths of the lightcurves in the bottom panel. Bottom panel: lightcurves at different wavelengths of the \lya\ line during the 28 October 2003 flare. The time profiles of the neighbouring Si~III line is also shown as a grey histogram. The vertical dashed and solid lines denote the approximate times of the pre-flare and flare spectra in the top panel, respectively. The GOES SXR lightcurve (black) is also shown for context.}
\label{sorce_lya}
\end{center}
\end{figure}

\section{SORCE/SOLSTICE Flare Observations}
\label{sec:sorce_lya}
The SOLSTICE instrument \citep{mccl05a,mccl05b onboard the SORCE satellite} typically takes spectral irradiance measurements across the EUV range once per orbit. However, during the X17 flare that occurred on 28 October 2003 (SOL2003-10-28T11:10) it was fortuitously scanning through the \lya\ line in steps of 0.35\AA\ at $\sim$1~minute cadence ($\sim$1~second exposure per step). \cite{wood04} reported only a 20\% increase in the core of the line, but the wings of the line were found to increase by a factor of two. This is significantly higher than the 6\% increase reported by \cite{brek96} during the end of the impulsive phase of the X3 flare on 27 February 1992 using the SOLSTICE instrument on the Upper Atmosphere Research Satellite. The line profiles from before and during the 28 October 2003 flare are shown in the top panel of Figure~\ref{sorce_lya}. The coloured lightcurves in the bottom panel are from the wavelength bins denoted by the same vertical coloured lines in the top panel. 

In agreement with \cite{wood04}, the time profiles for each wavelength bin peaked during the rise in SXR, confirming an impulsively heated atmosphere and the blue wing enhancement was found to be greater than that of the red wing, although this may be due to the presence of a blended emission line around 1211\AA\ (possibly S~X; log~T=6.2). The larger enhancements in both wings relative to the core are most likely due to opacity effects as photons generated in the core of the line get redistributed (see \citealt{wood95}). Generating a separate lightcurve for the nearby Si~III line at 1206.5\AA\ (log~T=4.7), shows an impulsive behaviour peaking in the rise phase of the SXR (grey histogram; bottom panel of Figure~\ref{sorce_lya}). Therefore, all emission within the $\sim$22\AA\ wide wavelength range observed by SOLSTICE is clearly indicative of a rapidly-responding chromosphere to a sudden deposition of energy. This implies that neither opacity effects nor the nearby Si~III line are responsible for the peculiar behaviour of the MEGS-P data.

\section{GOES/EUVS-E Flare Observations}
\label{sec:goes_lya}
The three most recent GOES Satellites, GOES-13, -14, and, -15, each carry onboard an EUV Sensor\footnote{\url{http://www.ngdc.noaa.gov/stp/satellite/goes/doc/GOES_NOP_EUV_readme.pdf}} (EUVS; \citealt{vier07,evan10}) in addition to the more familiar X-Ray Sensor. GOES-13 has been taking sporadic measurements in the EUV since mid-2006, while GOES-14 data are available from mid-2009 to mid-2010, and again for a few months in late 2012. GOES-15 has been much more consistent, observing the Sun continuously since early 2010. The A and B channels on each EUVS instrument are centered around 100\AA\ and 304\AA\ (the resonance line of He~II), respectively, while the E channel has a width of $\sim$100\AA\ (1180--1270\AA) spanning the \lya\ line in a broadband manner similar to MEGS-P. The calibrated GOES/EUVS Version 3 data are available at 1~minute cadence, while the raw data are available every 10 seconds.

\begin{figure}[!t]
\begin{center}
\includegraphics[width=0.49\textwidth]{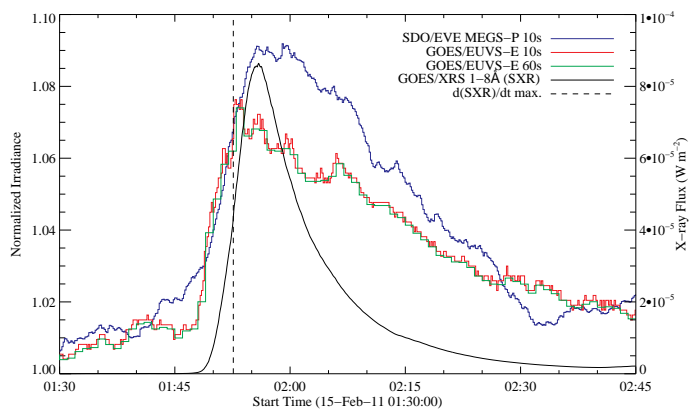}
\caption{A comparison between normalised irradiance as observed by SDO/EVE MEGS-P (blue curve) and GOES/EUVS-E at 10s (red curve) and 60s (green curve) cadence during the 15 February 2011 flare. The GOES SXR light curve is also plotted in black for context. The vertical dashed line denotes the time of the peak of the GOES SXR derivative.}
\label{goes_eve_lya}
\end{center}
\end{figure}

Fortunately, GOES-15 also observed the 15 February 2011 flare. The normalised lightcurves from both GOES/EUVS-E (red and green curves) and EVE MEGS-P (blue curve) are plotted in Figure~\ref{goes_eve_lya}, along with the SXR lightcurve (black) for context. There are obvious discrepancies between the changes in relative intensity quoted by each instrument, which can most likely be attributed to calibration issues. However, the differences in temporal behaviour are much more intriguing. The MEGS-P curve shows a distinct `pre-flare' increase beginning at $\sim$01:40~UT, and continues to increase over the subsequent $\sim$20~minutes, peaking close to - or possibly after - the GOES SXR peak. The EUVS-E emission, on the other hand, increases abruptly around 01:48~UT and peaks around 5 minutes later, close to the peak of the GOES SXR derivative (01:52:36~UT; vertical dashed line). The EUVS-E emission then begins to decay while MEGS-P emission continues to rise. Fluxes from both instruments return to their pre-flare values around an hour later. Therefore, despite both GOES/EUVS-E and SDO/EVE MEGS-P being full-Sun, broadband, high-cadence instruments centered on the \lya\ line, they exhibit distinctly different time profiles for the same event.

\section{Summary and Conclusions}
\label{sec:conc}
In the 5 years since its launch, SDO/EVE has detected increased \lya\ emission in dozens of flares over a range of magnitudes at 10~s cadence with its MEGS-P broadband diode. In all cases, the \lya\ time profile exhibits a slowly-varying, gradual behaviour, similar to that seen by thermal plasma emitting in SXR. By contrast, LyC and other intrinsically chromospheric emission appears much more impulsive, mimicking that of the HXR emission, as would be expected from an electron-beam-heated atmosphere. In comparison, spectrally- and temporally-resolved \lya\ emission from SOLSTICE appears impulsive during at least one previously observed event. the \lya\ emission detected by GOES/EUVS-E also appears to peak during the impulsive phase of the 15 February 2011 flare; distinctly different from that displayed by EVE which peaks some 5--10 minutes later. This suggests that the broadband nature of the EVE (and LYRA) measurements are in some way ``smoothing out'' this intrinsic, bursty nature. Perhaps other, higher temperature lines or continua are contributing to the passband, or a data processing algorithm in the EVE pipeline is responsible. While these subtle differences may not significantly impact research into effects on changes in the solar irradiance on the upper terrestrial atmosphere, they could dramatically affect how we interpret such changes in \lya\ in the context of solar flares themselves. 

One way to resolve this issue would be to convolve synthetic spectra from the RADYN radiative hydrodynamic code \citep{allr05,allr15} with the broadband response functions to establish what other lines and continua might be contributing to the passbands during large events. Spectrally-resolved \lya\ flare observations from other instruments are not suited to this task for a number of reasons: SORCE/SOLSTICE data only span 22\AA\ while the MEGS-P passband is 100\AA\ wide; SOHO/SUMER never observed solar flares due to the sensitivity of its optics; and Skylab data were often saturated during large events. The spectral response of the MEGS-P diode could also be mapped by calibrating the EVE rocket instrument using a monochrometer with a deuterium lamp. Once this issue is resolved, simultaneous \lya\ and LyC observations from EVE (along with \lyb, \lyc, etc., and other chromospheric diagnostics) will be a valuable tool for investigating heating of the lower solar atmosphere. Understanding the broadband nature of \lya\ observations from EVE will also help prepare for the influx of flare data from future instruments. For example, the Mars Atmosphere Variation and Evolution (MAVEN; \citealt{epar15}) features a \lya\ radiometer (1210--1220\AA) for measuring the effects of changes in solar EUV irradiance on the martian atmosphere. The Solar Orbiter mission - due for launch in 2018 - also includes a \lya\ channel as part of its Extreme Ultraviolet Imager instrument.

\begin{acknowledgements}
The authors would like to thank the anonymous referee for their constructive comments that greatly enhanced this paper. They also thank Marty Snow (LASP) for access to the high-cadence SOLSTICE data, Janet Machol (NOAA) for help with the GOES EUVS-E data, and Mihalis Mathioudakis (QUB) for many stimulating discussions on this issue. This work was supported by NASA LWS/TR\&T grant NNX11AQ53G and LWS/SDO Data Analysis grant NNX14AE07G.
\end{acknowledgements}

\end{document}